\documentstyle[aps,epsf,twocolumn,feynmf]{revtex}
\def\be{\begin{eqnarray}}
\def\ee{\end{eqnarray}}
\def\no{\nonumber}

\def\la{\langle}
\def\ra{\rangle}

\newcommand{\tr}{{\rm tr}}

\renewcommand{\th}{\theta}
\newcommand{\Sg}{\Sigma}
\newcommand{\dl}{\delta}

\newcommand{\arr}[4]{
\left(\begin{array}{cc}
#1&#2\\
#3&#4
\end{array}\right)
}

\newcommand{\eq}{\begin{equation}}
\newcommand{\eqx}{\end{equation}}
\newcommand{\eqn}{\begin{eqnarray}}
\newcommand{\eqnx}{\end{eqnarray}}

\newcommand{\f}[2]{\frac{#1}{#2}}
\newcommand{\GG}{{\cal G}}
\newcommand{\DD}{{\cal D}}

\newcommand{\One}{\mbox{\bf 1}}

\newcommand{\cor}[1]{\la{#1}\ra}

\newcommand{\br}[1]{\overline{#1}}

\newcommand{\psib}{\br{\psi}}

\newcommand{\lm}{\lambda}

\newcommand{\chis}{\chi_*}
\newcommand{\gmv}{\gamma_5}

\begin{document}

\draft
\title{\bf  The $U(1)$ Problem in Chiral Random Matrix Models}

\author{{\bf Romuald A. Janik}$^{1}$,
{\bf Maciej A.  Nowak}$^{2}$, {\bf G\'abor Papp}$^{3}$,
 and  {\bf Ismail Zahed}$^4$} 

\address{$^1$ Department of Physics,
Jagellonian University, 30-059 Krakow, Poland\\ 
$^2$ GSI, Plankstr.1, D-64291 Darmstadt \& Institut f\"{u}r Kernphysik,
TH Darmstadt, Germany \& \\Department of Physics,
Jagellonian University, 30-059 Krakow, Poland;\\
$^3$GSI, Plankstr. 1, D-64291 Darmstadt, Germany \&\\
Institute for Theoretical Physics, E\"{o}tv\"{o}s University,
Budapest, Hungary\\
$^4$Department of Physics, SUNY, Stony Brook, New York 11794, USA.}
\date{\today}
\maketitle

\begin{abstract}
We show that conventional asymmetric chiral random matrix models (ChRMM),
with a gaussian distribution in the asymmetry,
provide for a screening of the topological charge and 
a resolution of the $U(1)$ problem in the unquenched approximation. 
Our exact results to order $1/N$ are in agreement with numerical estimates
using large ensembles of asymmetric ChRMM with gaussian distributions.
\end{abstract}

\setlength{\unitlength}{1mm}
\begin{fmffile}{ng}

\fmfcmd{%
vardef polycross expr n =
  save ap, i;
  path ap;
  ap = fullcircle;
  for i = 1 upto n:
    origin -- .5 dir (360(i-.5)/n) --
  endfor
  (ap rotated (360(n-.5)/n)) --
  cycle
enddef;
}
\fmfwizard

\def\gluon{%
  \begin{fmfgraph}(6,10)%
	\fmftop{i1,u1,u2,o1}%
	\fmfbottom{i2,l1,l2,o2}%
	\fmffreeze%
    \fmfforce{(0,h)}{i1}%
    \fmfforce{(.35w,h)}{u1}%
    \fmfforce{(.65w,h)}{u2}%
    \fmfforce{(w,h)}{o1}%
    \fmfforce{(0,0)}{i2}%
    \fmfforce{(.35w,0)}{l1}%
    \fmfforce{(.65w,0)}{l2}%
    \fmfforce{(w,0)}{o2}%
	\fmf{plain}{i1,u1,l1,i2}%
	\fmf{plain}{o1,u2,l2,o2}%
  \end{fmfgraph}%
}
\def\dgluon{%
  \begin{fmfgraph}(6,10)%
	\fmftop{i1}%
	\fmfbottom{i2}%
	\fmfforce{(.5w,h)}{i1}%
	\fmfforce{(.5w,0)}{i2}%
	\fmf{zigzag}{i1,i2}%
  \end{fmfgraph}%
}
\def\dglDS{%
  \begin{fmfgraph}(10,10)%
	\fmftop{i1,ur1,ur2,o1}%
	\fmfbottom{i2,lr1,lr2,o2}%
	\fmffreeze%
    \fmfforce{(0,h)}{i1}%
    \fmfforce{(.7w,h)}{ur1}%
    \fmfforce{(.9w,h)}{ur2}%
    \fmfforce{(w,h)}{o1}%
    \fmfforce{(0,0)}{i2}%
    \fmfforce{(.7w,0)}{lr1}%
    \fmfforce{(.9w,0)}{lr2}%
    \fmfforce{(w,0)}{o2}%
	\fmffreeze%
	\fmf{zigzag}{i1,i2}%
	\fmf{plain}{o1,ur2,lr2,o2}%
	\fmf{plain}{i1,vu,ur1,lr1,vl,i2}%
	\fmfv{d.sh=circle,d.f=1,d.si=0.25w}{vu}%
	\fmfv{d.sh=circle,d.f=1,d.si=0.25w}{vl}%
  \end{fmfgraph}%
}
\def\gamone{%
   \begin{fmfgraph}(15,8)%
	\fmfcurved%
	\fmfsurround{o,b,i,t}%
	\fmf{plain,right=.25}{i,t,o}%
	\fmf{plain,left=.25}{i,b,o}%
	\fmfv{d.sh=circle,d.f=1,d.si=.17w}{t}%
	\fmfv{d.sh=circle,d.f=1,d.si=.17w}{b}%
	\fmfv{d.sh=polycross 4,d.si=.25w,d.f=0}{o}%
	\fmfv{d.sh=polycross 4,d.si=.25w,d.f=0}{i}%
   \end{fmfgraph}%
}
\def\gamtwo{%
   \begin{fmfgraph}(20,8)%
	\fmftop{i,ul,vu,ur,o}%
	\fmfbottom{i1,ll,vl,lr,o1}%
	\fmfforce{(0,.5h)}{i}%
	\fmfforce{(w,.5h)}{o}%
	\fmffreeze%
	\fmf{plain,left=.25}{i,ul}%
	\fmf{plain,right=.25}{i,ll}%
	\fmf{plain}{ul,ur}%
	\fmf{plain}{ll,lr}%
	\fmf{plain,left=.25}{ur,o}%
	\fmf{plain,right=.25}{lr,o}%
	\fmf{zigzag}{vu,vl}%
	\fmfv{d.sh=circle,d.f=1,d.si=.13w}{ul}%
	\fmfv{d.sh=circle,d.f=1,d.si=.13w}{ur}%
	\fmfv{d.sh=circle,d.f=1,d.si=.13w}{ll}%
	\fmfv{d.sh=circle,d.f=1,d.si=.13w}{lr}%
	\fmfv{d.sh=polycross 4,d.si=.2w,d.f=0}{o}%
	\fmfv{d.sh=polycross 4,d.si=.2w,d.f=0}{i}%
   \end{fmfgraph}%
}

\fmfset{dot_len}{1mm}

\section{Introduction}

In QCD the axial $U(1)$ symmetry is not apparent in the spectrum, 
although a non-vanishing quark condensate suggests the existence of a ninth 
Goldstone boson. This is the $U(1)$ problem. The resolution of this apparent
problem is believed to follow from the chiral anomaly \cite{ADLER}. 't Hooft
has suggested a specific mechanism using instantons \cite{THOOFT}. Witten has 
proposed a resolution in the context of the large $N_c$ (number of colors) 
limit \cite{WITTEN}, an idea that was interpreted by Veneziano in terms of 
vector ghosts \cite{VENEZIANO}.

In all existing scenarios for the 
resolution of the $U(1)$ problem, it is crucial that the 
topological susceptibility is nonzero in both the quenched and unquenched
(with massive quarks) approximation. Lattice simulations appear to
support this assumption both for quenched \cite{LATTICE1} and unquenched 
\cite{LATTICE2}, although the idea may be at odd with translational 
invariance and current identities \cite{YAZA}. This notwithstanding, it was 
argued by few that a statistical ensemble of topological charges may yield 
to the screening of the bare topological susceptibility and the resolution 
of the $U(1)$ problem \cite{VERGELES}. This 
scenario will be considered in the context of standard ChRMM 
\cite{MICRO,MICROMORE,SOME,US}.

Standard ChRMM follow from the constant mode sector of the instanton liquid 
model \cite{SHURYAK,DIAKONOV,USBOOK}, regarded as a statistical ensemble of topological zero 
modes.  As models, 
they offer a minimal framework for discussing the interplay between 
(chiral) symmetry, typical scales and the thermodynamical limit. In these 
models gaussian randomness is enough to cause the spontaneous breaking of 
chiral symmetry (hermitean matrices) or the spontaneous breaking of 
holomorphic symmetry (nonhermitean matrices) in the thermodynamical limit. 
They have also been used to investigate issues related to the universality 
of ``noise-fluctuations" in Dirac spectra both in the microscopic 
\cite{MICRO} and macroscopic \cite{MACRO} limit.

In so far, most of the analyses carried in the context of standard ChRMM 
have been conducted using standard and symmetric ChRMM. In the context of 
the instanton liquid model this means that the number of instantons and
the number of antiinstantons is fixed. In this case there is no resolution 
of the $U(1)$ problem. In this letter, we will consider a statistical ensemble
of asymmetric ChRMM with a gaussian distribution for the asymmetry. The
physical motivation for this model stems from the coarse grained version of
the instanton liquid model \cite{ALKOFER}. In section 2, we will
discuss basic QCD Ward identities emphasizing the role of a non-vanishing 
topological susceptibility in the resolution of the $U(1)$ problem.
In section 3, we streamline the phenomenological arguments for standard but 
asymmetric ChRMM
with a gaussian distribution for the matrix asymmetry.  In section 4, we use
a diagrammatic analysis based on a $1/N$ expansion to solve the $U(1)$ 
problem. Our analytical results for the unquenched topological, scalar and
pseudoscalar susceptibilities are in agreement with numerical estimates using
large asymmetric matrices. Our conclusions are in section~5. 
We elaborate on some derivations in the Appendices. Throughout, we 
will use four-dimensional arguments in Minkowski space. The transcription to
ChRMM will be done through their Euclidean four-dimensional analogue.

\section{QCD Ward Identity}

The topological term in the QCD Lagrangian $\theta\,\,\Xi$ is a total 
divergence
\be
\Xi = \frac {g^2}{32\pi^2} G_{\mu\nu}^a\tilde{G}^{\mu\nu a}=\partial^{\mu} 
K_{\mu}
\label{1}
\ee
where $K_{\mu}$ is the Loos-Chern-Simons current. For $N_f$ flavors of quarks
with current masses $m_f$, the gauge-invariant flavor axial-singlet current
$j_{\mu 5}^{\rm inv}$ is anomalous,
\be
\partial^{\mu} j_{\mu 5}^{\rm inv} = 2N_f \Xi + 2 \sum_f^{N_f} m_f 
\overline{\psi}_f i\gamma_5 \psi_f
\label{2}
\ee
In a $\theta$ state, the expectation value of (\ref{2}) implies
\be
0= 2N_f \langle\theta | \Xi |\theta \rangle + 2 \sum_f^{N_f}
\langle\theta | m_f 
\overline{\psi}_f i\gamma_5 \psi_f |\theta\rangle
\label{3}
\ee
by translational invariance.
In particular, the topological susceptibility $\chi (\theta )$ is given by
\be
\chi (\theta ) = \frac {\partial\langle\theta |\Xi 
|\theta\rangle}{\partial\theta} = -\frac 1{N_f} \sum_f^{N_f} 
\frac {\partial}{\partial\theta} \langle\theta |
m_f  \overline{\psi}_f i\gamma_5 \psi_f |\theta\rangle 
\label{4}
\ee

The absence of a physical massless $U(1)$ boson for quarks of equal
masses $m_f=m$, gives \cite{YAZA}
\be
0=&& \int d^4x \partial^{\mu} \langle \theta | T^* j_{\mu 5}^{\rm inv} (x) 
\overline{\psi}i\gamma_5 \psi (0)|\theta \rangle\nonumber\\
=&&-2i\langle\theta | \overline{\psi}\psi |\theta \rangle
-2i\frac {N_f^2}{m} \chi (\theta )\nonumber\\
&& + 2 m\int d^4x \langle\theta |T^*
\overline{\psi} i\gamma_5 \psi (x)\,
\overline{\psi} i\gamma_5 \psi (0) |\theta\rangle
\label{5}
\ee
where we have made use of (\ref{3},\ref{4}). $T^*$ is the covariantized 
T-product. For $\theta =0$, it follows from (\ref{5}) that
\be
i\chi_{top} = -\frac {im}{N_f^2} \langle\overline{\psi}\psi \rangle
+\frac {m^2}{N_f^2}\int d^4x \langle  T^*\overline{\psi} i\gamma_5\psi (x)\,\,
\overline{\psi} i\gamma_5\psi (0) \rangle
\label{6}
\ee
where we have set $\chi_{top}=\chi (0)$, in agreement with earlier results
\cite{CREWTHER,SHIFMAN}.
The resolution of the $U(1)$ problem stems from the observation that
for small $m$, the absence of a $U(1)$ Goldstone mode requires that 
$\chi_{top} =-m\langle\overline{\psi}\psi\rangle /N_f^2$ to order ${\cal O}
(m^2)$.

A rerun of the above argument for the $SU(N_f)$ currents and densities
yields the flavor non-singlet relation
\be
0=&&-\frac i2 \langle \overline{\psi} \,\,[\lambda_I, \lambda_J
]_+\psi \rangle \nonumber\\
&&+ m \int d^4x
\langle  T^*\overline{\psi} i\gamma_5\lambda_I\psi (x)\,\,
\overline{\psi} i\gamma_5\lambda_J\psi (0)  \rangle
\label{7}
\ee
which shows that for small $m$ there should be a multiplet of $N_f^2-1$
Goldstone modes. Contrasting (\ref{6}) with (\ref{7}) shows the importance 
of a non-vanishing topological susceptibility in the resolution of the $U(1)$
problem. It was originally argued by Witten \cite{WITTEN} that further
consistency
with large $N_c$ arguments requires that the quenched topological susceptibility 
$\chi_*$ be nonzero as well. Current lattice simulations seem to support this
conjecture \cite{LATTICE1},  although there may be subtleties as we indicated 
above \cite{YAZA}.

\section{ChRMM}

A number of effective models aimed at describing the long wavelength physics of 
the QCD vacuum including the $U(1)$ anomaly, have been put forward by 
several authors~\cite{SCHECHTER}. In these effective
models it is important that point-like pseudoscalars are coupled to point-like 
glueballs to achieve consistency with the QCD anomalies (axial and 
scale anomaly). 

Such effective models arise naturally from 
microscopic descriptions of the QCD vacuum using a random ensemble of 
instantons and antiinstantons \cite{SHURYAK,DIAKONOV,USBOOK}. In the
``coarse grained instanton model'' \cite{USBOOK,ALKOFER},
the effective ``glueball" fields are identified with
 \eqn
    \frac{g^2}{32\pi^2} G \cdot G(x) &&\rightarrow (n_++n_-)(x) \\
    \frac{g^2}{32\pi^2} G \cdot \tilde{G}(x)&&\rightarrow (n_+-n_-)(x)
 \label{ident}
\eqnx
where $n_{\pm} (x)$ is the density of instantons (+) and antiinstantons ($-$).
The $U(1)$ anomaly is saturated by assuming that the distribution in 
the number difference $(n_+-n_-)$ (susceptibility of the quenched vacuum) is 
gaussian. The scale anomaly is also saturated by assuming that 
the distribution in the number density $(n_++n_-)$ (compressibility of 
the quenched vacuum) follows from a logarithmic ensemble 
\cite{ALKOFER,SCHECHTER}.

\subsection{Symmetric ChRMM}

Standard ChRMM are schematic versions of current descriptions of the 
instanton liquid model \cite{SHURYAK,DIAKONOV,USBOOK}. One 
essentially truncates the QCD partition function to the space of 
instanton-antiinstanton zero modes and furthermore assumes that the 
fermionic overlaps are randomly distributed with a Gaussian measure. 
Specifically
\eq
\label{e.zsquare}
Z=\left\la \mbox{det}^{N_f} {\bf Q} \right\ra_A
\eqx
where 
\eq
{\bf Q}=\arr{im_f}{A^{\dagger}}{A}{im_f}
\eqx
and $A$ is an $N \times N$ matrix distributed with a weight
\eq
P(A)=e^{-N\Sg \tr AA^{\dagger}}
\eqx
where $\Sg$ is a fixed scale related to the quark condensate through
$<\psi^{\dagger}\psi >=1/\pi \Sigma$. For simplicity
$\Sigma=1$ in the arguments to follow. The large $N$ limit is directly 
linked to the thermodynamic limit through the choice $N/V_4 =1$.
In other words the number of zero modes is made commensurate to the
four volume $V_4$, to insure a non-vanishing chiral condensate in the 
thermodynamical limit.

One can easily check that for such ensembles, the topological susceptibility
is zero. This is due to the fact that in (\ref{e.zsquare}) we have used
symmetric matrices, {\it i.e.} in the instanton liquid model,
we have set the number of instantons equal to the number of antiinstantons.

\subsection{Asymmetric ChRMM}

The above constraint can be relaxed following \cite{ALKOFER,ZAHED,DIAKONOVMORE}
through
\eq
\label{e.zasym}
Z=\sum^*_{n_{\pm}} Z_{n_+,n_-} e^{-\f{\chi^2}{2\chis N}}
\eqx
where $\chi=n_+-n_-$ is the difference between the number of instantons
and antiinstantons and $Z_{n_+,n_-}$ is as before (\ref{e.zsquare}), but
with the $A$'s now rectangular and $n_+\times n_-$ complex valued.
The star in (\ref{e.zasym}) indicates that the sum is restricted to
$n_++n_-=2N$ with $N$ eventually going to infinity. The issue related to
the scale anomaly \cite{SHIFMAN1} (for our case an unrestricted sum in 
(\ref{e.zasym})) will be discussed elsewhere.
The topological susceptibility in the limit of infinite quark mass
(quenched) is in this case just
\eq
\chi_{top}=\f{1}{N}\cor{\chi\chi}=\chis
\eqx
Although (\ref{e.zasym}) describes a statistical ensemble of
point-like topological defects, that is zero correlation length,
the latters are spread over a gaussian of width $\sqrt{N\chi_*}$, 
resulting into a finite correlation length or topological susceptibility
in the thermodynamical limit.
In the next section we evaluate the pseudoscalar and topological
susceptibilities for the ensemble (\ref{e.zasym}) in the context of
a $1/N$ expansion.

\section{$U(1)$ Problem in ChRMM}

A measure of the axial-singlet charge in the QCD vacuum in the presence of 
light quarks is given by the pseudoscalar axial-singlet susceptibility
\eq
\chi_{ps}=\f{1}{V_4}\int d^4x\cor{T^*\psib\gmv\psi(x)\,\psib\gmv\psi(0)}
\label{50}
\eqx
A resolution of the $U(1)$ problem requires that $\chi_{ps}$ remains finite
as $V_4\rightarrow\infty$ followed by $m\rightarrow 0$.

To help investigate (\ref{50}) in the context of standard ChRMM, we recast 
(\ref{e.zasym}) in the form of a partition function over Grassmann variables 
\be
Z=\sum^*_{n_{\pm}}\int d\psi^{\dagger} d\psi dA e^{-{\sum_{N_f}
	\psi^{\dagger}{\bf Q}\psi -N\tr AA^{\dagger} 
-\chi^2/ 2N\chi_*}}
\label{51}
\ee
The Euclidean analogue of (\ref{50}) in ChRMM is
\eqn
\chi_{ps}&=&\f{1}{N}\left[-\cor{\tr {\bf Q}^{-1}\gmv {\bf Q}^{-1}\gmv} 
\right] \label{TR}\nonumber\\
 &+&\f{1}{N}\left[\cor{\tr ({\bf Q}^{-1}
\gmv) \tr({\bf Q}^{-1}\gmv)}_c\right]
\label{TRTR}
\eqnx
where all averages are carried using (\ref{51}) that is (\ref{e.zasym}).
A naive expectation based upon large $N$ counting rules suggests that
the second term in (\ref{TRTR}) is subleading in the thermodynamic limit.
As we will show below, this expectation is not born out by calculations.

\subsection{The Rules and Resolvent}

\vskip .2cm
{$\bullet \,\,$}{\it Feynman Rules}
\vskip .1cm

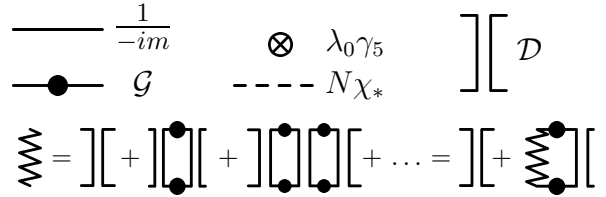
\begin{figure}[htbp]
\begin{tabular}{cccc}
%
%
\parbox{12mm}{%
  \begin{fmfgraph}(12,3)
	\fmfleft{i1}
	\fmfright{o1}
	\fmf{plain}{o1,i1}
  \end{fmfgraph}
} & \raisebox{2mm}{\Large $\frac{1}{-im}$} \hspace*{4mm}
&
%
%
\phantom{spa}
\parbox{6mm}{%
  \begin{fmfgraph}(3,3)
	\fmfright{v}
	\fmfforce{(.5w,.5h)}{v}
	\fmfv{d.sh=polycross 4,d.si=w,d.f=0}{v}
  \end{fmfgraph}
} & \mbox{\large $\lambda_0 \gamma_5$} \\
%
%
\parbox{12mm}{\raisebox{-2mm}{%
  \begin{fmfgraph}(12,3)
	\fmfleft{i1}
	\fmfright{o1}
	\fmf{plain}{o1,v,i1}
	\fmfv{d.sh=circle,d.f=1,d.si=0.2w}{v}
  \end{fmfgraph}
}} & \mbox{\large $\cal{G}$} \hspace*{4mm}
&
%
%
\parbox{12mm}{\raisebox{-3mm}{%
  \begin{fmfgraph}(12,3)
	\fmfleft{i1}
	\fmfright{o1}
	\fmf{dashes}{o1,i1}
  \end{fmfgraph}
}} & \mbox{\large $N \chi_*$}
\end{tabular}\phantom{space}
\begin{tabular}{cc}
%
%
\parbox{6mm}{%
	\gluon
} & \mbox{\large ${\cal D}$} \\
\end{tabular}\\[3mm]

\setlength{\unitlength}{0.75mm}
%
%
\parbox{4mm}{\dgluon} =
\parbox{4mm}{\gluon} +
\parbox{8mm}{\raisebox{-2mm}{%
  \begin{fmfgraph}(10,10)
	\fmftop{i1,ul1,ul2,ur1,ur2,o1}
	\fmfbottom{i2,ll1,ll2,lr1,lr2,o2}
	\fmffreeze
    \fmfforce{(0,h)}{i1}
    \fmfforce{(.1w,h)}{ul1}
    \fmfforce{(.25w,h)}{ul2}
    \fmfforce{(.75w,h)}{ur1}
    \fmfforce{(.9w,h)}{ur2}
    \fmfforce{(w,h)}{o1}
    \fmfforce{(0,0)}{i2}
    \fmfforce{(.1w,0)}{ll1}
    \fmfforce{(.25w,0)}{ll2}
    \fmfforce{(.75w,0)}{lr1}
    \fmfforce{(.9w,0)}{lr2}
    \fmfforce{(w,0)}{o2}
	\fmf{plain}{i1,ul1,ll1,i2}
	\fmf{plain}{o1,ur2,lr2,o2}
	\fmf{plain}{ul2,vu,ur1,lr1,vl,ll2,ul2}
	\fmfv{d.sh=circle,d.f=1,d.si=0.25w}{vu}
	\fmfv{d.sh=circle,d.f=1,d.si=0.25w}{vl}
  \end{fmfgraph}
}}  + 
\parbox{14mm}{\raisebox{-2mm}{%
  \begin{fmfgraph}(20,10)
	\fmftop{i1,ul1,ul2,um1,um2,ur1,ur2,o1}
	\fmfbottom{i2,ll1,ll2,lm1,lm2,lr1,lr2,o2}
	\fmffreeze
    \fmfforce{(0,h)}{i1}
    \fmfforce{(.1w,h)}{ul1}
    \fmfforce{(.2w,h)}{ul2}
    \fmfforce{(.45w,h)}{um1}
    \fmfforce{(.55w,h)}{um2}
    \fmfforce{(.8w,h)}{ur1}
    \fmfforce{(.9w,h)}{ur2}
    \fmfforce{(w,h)}{o1}
    \fmfforce{(0,0)}{i2}
    \fmfforce{(.1w,0)}{ll1}
    \fmfforce{(.2w,0)}{ll2}
    \fmfforce{(.45w,0)}{lm1}
    \fmfforce{(.55w,0)}{lm2}
    \fmfforce{(.8w,0)}{lr1}
    \fmfforce{(.9w,0)}{lr2}
    \fmfforce{(w,0)}{o2}
	\fmf{plain}{i1,ul1,ll1,i2}
	\fmf{plain}{o1,ur2,lr2,o2}
	\fmf{plain}{ul2,vu1,um1,lm1,vl1,ll2,ul2}
	\fmf{plain}{ur1,vu2,um2,lm2,vl2,lr1,ur1}
	\fmfv{d.sh=circle,d.f=1,d.si=0.1w}{vu1}
	\fmfv{d.sh=circle,d.f=1,d.si=0.1w}{vl1}
	\fmfv{d.sh=circle,d.f=1,d.si=0.1w}{vu2}
	\fmfv{d.sh=circle,d.f=1,d.si=0.1w}{vl2}
  \end{fmfgraph}
}}  + $\ldots$ =
\parbox{3mm}{%
	\gluon
}  + \hspace*{1mm}
\parbox{4mm}{\dglDS}\\[1mm] 
\setlength{\unitlength}{1mm}
\caption{Feynman rules. In the upper part: the bare quark propagator
$1/(-im)$, the dressed one (${\cal G}$), the $\gamma_5$ vertex, the bare
$\chi$ propagator and the bare gluon propagator ${\cal D}$. In the lower
part the dressing of the gluon propagator is shown, the last equality
representing the Schwinger-Dyson equation.} 
\label{fig-rules}
\end{figure}

To assess (\ref{TRTR}) we use a $1/N$ expansion. The Feynman rules 
associated to (\ref{50}) are shown in Fig.~\ref{fig-rules}. The bare 
quark propagator is $1/(- im)$  and the bare gluon propagator is
\eqn
\DD=\cor{A^a_b A^c_d}&=&\left[\f{1}{N}\arr{1}{0}{0}{0}_{ad} \otimes 
\arr{0}{0}{0}{1}_{bc} \right.  \nonumber \\  &+&
\left. \f{1}{N}\arr{0}{0}{0}{1}_{ad} \otimes \arr{1}{0}{0}{0}_{bc}\right]
\otimes {\cal F}
\eqnx
Here $a$, $b$, $c$ and $d$ run from 1 to $2N$, $\otimes$ denotes tensor
product and ${\cal F}$ is a flavor 
bearing matrix with flavor indices $f_i$. For $N_f=3$
\eq
{\cal F}^{f_1 f_3}_{f_2 f_4}=
\dl^{f_1}_{f_4} \dl^{f_2}_{f_3}=
\f{1}{2}([\lm_0]^{f_1}_{f_2}[\lm_0]^{f_3}_{f_4}+[\lm_I]^{f_1}_{f_2}[\lm_I]^{f_3}_{f_4})
\label{FID}
\eqx
following a standard  decomposition. In terms of (\ref{FID}), the bare gluon
propagator may be rewritten in the form
\eq
\DD=\f{1}{4N}\left((\One\lm_I) \otimes (\One\lm_I) - 
  (\gmv\lm_I)\otimes (\gmv\lm_I) \right)
\label{baregluon}
\eqx
with $I$ running from 0 to 8. We use the normalization 
$\tr \lm_I\lm_J=2\dl_{IJ}$. The flavor decomposition stemming from 
(\ref{baregluon}) is reminiscent of the flavor interaction in a
schematic quark-meson effective theory of the Nambu-Jona-Lasinio type.

\vspace*{-2mm}

\vskip .2cm
{$\bullet \,\,$}{\it Resolvent}
\vskip .1cm

For a single flavor we define the resolvent 
\eq
{\cal G}(z)=\frac{1}{2N}\left\la\frac{1}{z-\left(\begin{array}{cc}
	im & A^{\dagger}\\ 
	A  & im \end{array}\right)} \right\ra
\eqx
where averaging is now over rectangular matrices. For our purposes
we will need the resolvent calculated at $z=0$ only. For fixed $n_{\pm}$,
the result of the averaging over $A$ yields
\eq
\GG=g\One+g_5\gmv
\label{SPECIAL}
\eqx
with $\GG = \GG(0)$ and
\eqn
g&=&\f{\f{i}{2}\left(-m^2-2x^2+m\sqrt{4+m^2+\f{4x^2}{m^2}}\right)}{m(1-x^2)}
\nonumber\\
g_5&=&\f{\f{i}{2}x\left(m^2+2-m\sqrt{4+m^2+\f{4x^2}{m^2}}\right)}{m(1-x^2)}
\label{resss}
\eqnx
and
\eqn
x=\f{\chi}{2N}=\f{n_+-n_-}{2N}
\eqnx
Both ${\bf 1}$ and $\gamma_5$ in (\ref{SPECIAL}) are $2N\times 2N$ valued. 
The explicit form of the resolvent $g$ with $m=-iz$ in the chiral limit,
can be checked by other methods \cite{JNPWZ,FEINBERGZEE}\footnote{
Note that a different measure was used in~\cite{FEINBERGZEE}.}.
Both $g$ and $g_5$ are flavor additive. We observe that
\be
{\rm tr\,} {\cal G} = g+x g_5 = \frac{i}{2} \left(-m +\sqrt{4+
m^2+4\frac{x^2}{m^2}}\right)
\label{SPEC1}
\ee
and
\be
{\rm tr\,} \gamma_5 {\cal G} = x g + g_5 = \frac{ix}{m}
\label{SPEC2}
\ee 

For $x=0$ (square matrices) the discontinuity of (\ref{SPEC1}) 
\be
\nu_+ (\lambda , x=0 ) =&& -\frac 1{\pi} {\rm Im}\,\, {\rm tr\,} {\cal G} 
(m=-i\lambda + 0) 
\label{DISCONTINUITY}
\ee
is just Wigner's semicircle for the quark spectral distribution. The argument
of ${\cal G}$ in (\ref{DISCONTINUITY}) has been appended for clarity.
For $x\neq 0$ (asymmetric matrices) it is the 
resolvent in a configuration of unequal topological charges
\be
\nu_+ (\lambda ,x ) =  |x| \delta (\lambda ) +
\frac 1{2\pi |\lambda |} 
{\sqrt{(\lambda^2-\lambda_-^2)(\lambda_+^2-\lambda^2)}}
\ee
with $\lambda_{\pm}^2 =2\pm 2\sqrt{1-x^2}$. The delta function at the origin, 
reflects on the number of unpaired topological charges. Configurations with 
$x\neq 0$ do not break spontaneously chiral symmetry, owing to the occurrence
of a gap at the origin. The spontaneous breaking is triggered by the
neutral topological configurations with $x=0$.

We observe that the discontinuity in (\ref{SPEC2}) for fixed $x$
\be
\nu_- (\lambda, x) = -\frac{1}{\pi} {\rm Im}\,\, {\rm tr}(\gamma_5{\cal G}) 
(m=-i\lambda + 0) = x\delta (\lambda) 
\label{ati}
\ee
is a direct measurement of the difference in the 
spectral distribution between left-handed $\lambda_n^-$ and right handed 
$\lambda_n^+$ quark zero-modes. The result (\ref{ati}) reflects on the 
Atiyah-Singer index theorem. 

In the large $N$ limit, the fluctuations in $\chi$ are of order $\sqrt{N}$,
and a statistical averaging over an ensemble of asymmetric matrices
yields $x=0$ on the average.
A convenient way of interpreting the averaging over $\chi$
diagrammatically is to consider the partition function (\ref{e.zasym})
as describing the interaction of random matrices with a field $\chi$
whose propagator is $\cor{\chi\chi}=N\chis$ (see Fig.~\ref{fig-rules}).

\begin{figure}[htbp]
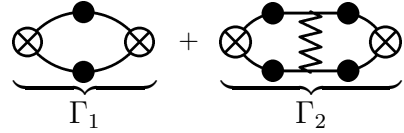

\be
\underbrace{\rule[-4mm]{2mm}{0mm}
	\parbox{17mm}{\gamone}}_{\mbox{\large $\Gamma_1$}} \ \ + \ \ 
\underbrace{\rule[-4mm]{2mm}{0mm}
	\parbox{22mm}{\gamtwo}}_{\mbox{\large $\Gamma_2$}} \no
\ee
\caption{`Trace' contribution ($\Gamma_1+\Gamma_2$) to the pseudoscalar 
susceptibility.}
\label{fig-ps}
\end{figure}

\vskip .2cm
{$\bullet \,\,$}{\it Gluon Ladder}
\vskip .1cm

In the next step we consider the resummation of the gluon ladder.
If the resolvent is diagonal\footnote{Throughout $m_u=m_d \neq m_s$.}  
\eq
\label{e.green}
\GG=(-i)(a_0\One\lm_0+a_8\One\lm_8)
\eqx
and the resummed planar propagator is decomposed as
\eq
\f{1}{4N}\left(  D^s_I \otimes (\One\lm_I) - 
 D^{ps}_I \otimes (\gmv\lm_I) \right)
\eqx
in analogy with (\ref{baregluon}),
then the resummation of the rainbow diagrams gives
\eqn
D^s_I&=&\One\lm_I+\f{1}{2}(\tr\lm_I\GG\lm_J\GG) D^s_J\\
D^{ps}_I&=&\gmv\lm_I-\f{1}{2}(\tr\lm_I\GG\lm_J\GG) D^{ps}_J
\eqnx
This resummation is shown in the form of a Schwinger-Dyson
equation in Fig~\ref{fig-rules}. The final result is
\eqn
\label{e.resum}
\DD_{resummed}&=&\f{1}{2N}\left( \left[\f{1}{2+d}\right]_{IJ}
\One\lm_I\otimes\One\lm_J  \right.  \nonumber \\       
   &-& \left.
\left[\f{1}{2-d}\right]_{IJ}\gmv\lm_I\otimes\gmv\lm_J\right)
\eqnx
where $d$ is $9\times 9$ matrix valued
\eqn
d_{IJ}=-{\rm tr} \lm_I \GG \lm_J \GG
\label{variousd}
\eqnx 
The denominators $2\pm d$
in the pseudoscalar channel can be interpreted within
the effective quark-meson analogy as generalized GMOR (Gell-Mann-Oakes-Renner) relations.
Using these results we may now evaluate the pseudoscalar susceptibility.

\subsection{The Puzzle}

Now consider the pseudoscalar axial-singlet susceptibility in ChRMM
\eq
\cor{\psi^{\dagger}\lm_0 \gmv\psi \psi^{\dagger} \lm_0 \gmv \psi}_c
\eqx
Let us first look at the diagrams corresponding to the 
first line (``trace term'') in (\ref{TR}). The graphs that contribute to the 
`trace' part are shown in Fig.~\ref{fig-ps}. The 
``trace-trace'' term, corresponding to second line 
of~(\ref{TRTR}), will be considered later.
We remark that main difference between the singlet and nonsinglet 
pseudoscalar correlators is the second line of~(\ref{TRTR}).

The first graph in Fig.~\ref{fig-ps}  is given by 
\eq
\Gamma_1=\tr\,\GG\gmv\lm_0 \GG \gmv\lm_0=\f{2}{3}\tr \,\GG^2
\eqx
The prefactor of $2/3$ follows from our choice of $\lambda_0=\sqrt{2/3}$.
Note that we may safely set $x=0$ here.
The second graph in Fig.~\ref{fig-ps} is given by
\eqn
\Gamma_2&=&\tr (\GG\gmv\lm_0\GG)\DD_{resummed}(\GG\gmv\lm_0\GG)
\eqnx
A simpler way to evaluate this graph is to use the identity

\centerline{%
\parbox{10mm}{\raisebox{-2mm}{%
  \begin{fmfgraph}(10,10)
	\fmftop{iu,u1,u2,ou}
	\fmfbottom{il,l1,l2,ol}
	\fmfforce{(0,h)}{iu}
	\fmfforce{(.25w,h)}{u1}
	\fmfforce{(.4w,h)}{u2}
	\fmfforce{(w,h)}{ou}
	\fmfforce{(0,0)}{il}
	\fmfforce{(.25w,0)}{l1}
	\fmfforce{(.4w,0)}{l2}
	\fmfforce{(w,0)}{ol}
	\fmf{plain}{iu,u1,l1,il}
	\fmf{plain}{ou,u2,l2,ol,v,ou}
	\fmfv{d.sh=polycross 4,d.si=.3w,d.f=0}{v}
  \end{fmfgraph}
}} \hspace*{3mm} = {\Large -}
\parbox{3mm}{\raisebox{-2mm}{%
  \begin{fmfgraph}(3,10)
	\fmftop{iu,ou}
	\fmfbottom{il,ol}
	\fmfforce{(0,h)}{iu}
	\fmfforce{(w,h)}{ou}
	\fmfforce{(0,0)}{il}
	\fmfforce{(w,0)}{ol}
	\fmf{plain}{iu,ou,v,ol,il}
	\fmfv{d.sh=polycross 4,d.si=w,d.f=0}{v}
  \end{fmfgraph}
}}
}
\vspace*{2mm}

Note that the bare ``quark lines'' carry only a factor 1 (amputated propagators).
This identity comes as a result of contracting a bare gluonic propagator 
(\ref{baregluon}) with a $\gamma_5 \lambda_0$ vertex.
Using the definition of the resummed propagator (Schwinger-Dyson equation)
we arrive at the useful identity

\vspace*{2mm}
\setlength{\unitlength}{0.9mm}
\centerline{%
\parbox{6mm}{\dgluon} {\bf --}
\parbox{6mm}{\gluon} = \ 
\parbox{10mm}{\dglDS}
	 {\Large $\Longrightarrow$} \hspace*{2mm}
\parbox{7mm}{\raisebox{-2mm}{%
  \begin{fmfgraph}(8,10)
	\fmfleft{i1,i2}
	\fmfright{o}
	\fmf{zigzag}{i1,i2}
	\fmf{plain,right=.25}{i1,vu,o}
	\fmf{plain,left=.25}{i2,vl,o}
	\fmffreeze
	\fmfshift{(0,-.13h)}{vu}
	\fmfshift{(0,.13h)}{vl}
	\fmfv{d.sh=circle,d.f=1,d.si=.3w}{vu}
	\fmfv{d.sh=circle,d.f=1,d.si=.3w}{vl}
	\fmfv{d.sh=polycross 4,d.si=.4w,d.f=0}{o}
  \end{fmfgraph}
}} \ = {\bf --} \hspace*{-2mm}
\parbox{6mm}{\raisebox{-2mm}{%
  \begin{fmfgraph}(6,10)
	\fmfleft{i1,i2}
	\fmfright{o}
	\fmf{plain,right=.25}{i1,o}
	\fmf{plain,left=.25}{i2,o}
	\fmfv{d.sh=polycross 4,d.si=.45w,d.f=0}{o}
  \end{fmfgraph}
}} \ {\bf --} \ 
\parbox{8mm}{\raisebox{-2mm}{%
  \begin{fmfgraph}(8,10)
	\fmfleft{i1,i2}
	\fmfright{o}
	\fmf{zigzag}{i1,i2}
	\fmf{plain,right=.25}{i1,o}
	\fmf{plain,left=.25}{i2,o}
	\fmfv{d.sh=polycross 4,d.si=.4w,d.f=0}{o}
  \end{fmfgraph}
}}
}
\vspace*{2mm}

Applying the above identity twice, yields

\vspace*{2mm}
\centerline{%
\begin{tabular}{rl}
\setlength{\unitlength}{0.75mm}
\parbox{17.3mm}{\gamtwo} = & {\bf --} \ \ \setlength{\unitlength}{0.65mm}
	\parbox{15mm}{\gamone} {\bf --} \ \ 
\parbox{12mm}{%
   \begin{fmfgraph}(17,8)%
	\fmftop{i,ul,vu,o}%
	\fmfbottom{i1,ll,vl,o1}%
	\fmfforce{(0,.5h)}{i}%
	\fmfforce{(w,.5h)}{o}%
	\fmffreeze%
	\fmf{plain,left=.25}{i,ul}%
	\fmf{plain,right=.25}{i,ll}%
	\fmf{plain}{ul,vu}%
	\fmf{plain}{ll,vl}%
	\fmf{plain,left=.25}{vu,o}%
	\fmf{plain,right=.25}{vl,o}%
	\fmf{zigzag}{vu,vl}%
	\fmfv{d.sh=circle,d.f=1,d.si=.15w}{ul}%
	\fmfv{d.sh=circle,d.f=1,d.si=.15w}{ll}%
	\fmfv{d.sh=polycross 4,d.si=.24w,d.f=0}{o}%
	\fmfv{d.sh=polycross 4,d.si=.24w,d.f=0}{i}%
   \end{fmfgraph}%
} \\[6mm]
 = & {\bf --} \ \parbox{15mm}{\gamone} {\bf +} \ 
\parbox{12mm}{%
   \begin{fmfgraph}(12,8)%
	\fmfleft{i}
	\fmfright{o}
	\fmf{plain,right=.5}{i,o}%
	\fmf{plain,left=.5}{i,o}%
	\fmfv{d.sh=polycross 4,d.si=.25w,d.f=0}{o}%
	\fmfv{d.sh=polycross 4,d.si=.25w,d.f=0}{i}%
   \end{fmfgraph}%
} \ {\bf +} \ 
\parbox{12mm}{%
   \begin{fmfgraph}(15,8)%
	\fmfcurved%
	\fmfsurround{o,b,i,t}%
	\fmf{plain,right=.25}{i,t,o}%
	\fmf{plain,left=.25}{i,b,o}%
	\fmf{zigzag}{t,b}
	\fmfv{d.sh=polycross 4,d.si=.25w,d.f=0}{o}%
	\fmfv{d.sh=polycross 4,d.si=.25w,d.f=0}{i}%
   \end{fmfgraph}%
}
\end{tabular}
}
\setlength{\unitlength}{1mm}
\vspace*{2mm}

We remark that in the sum $\Gamma_1+\Gamma_2$ the $\Gamma_1$ term cancels against
the first one in the above line and the result for the sum is simply the 
sum of the two remaining diagrams in the line above, that is
\eqn
\Gamma_1+\Gamma_2&=&+4N+(2N\cdot 2) 
\f{-1}{2N} \left[\f{1}{2-d}\right]_{00} (2N\cdot 2)\nonumber\\
&=&4N\left(1-2 \left[\f{1}{2-d}\right]_{00} \right)\nonumber\\
&=& \f{-4N}{3}\sum_{i=0}^{N_f}\f{2}{m_i(\sqrt{4+m_i^2}+m_i)}
\label{60}
\eqnx
The diagrams of Fig.~2 yield a Goldstone pole in 
the pseudoscalar axial-singlet correlator, hence the {\it a priori}
puzzle. 

Before we proceed to solve the puzzle, we note the remarkable cancelation
in the sum $\Gamma_1+\Gamma_2$. The self-energy insertions
on the two-quark lines and the one-gluon exchange diagram, conspire to 
remove the ``constituent" quark cut, leading to a Goldstone pole. For the
non-singlet channels this is the mechanism by which the Goldstone modes
emerge in a correlation function that does not abide by confinement. This
cancelation is reminiscent of the one taking place in two-dimensional QCD
in the planar approximation \cite{THOOFTTWO}. So it appears that the removal
of the ``constituent" quark cut is a simple consequence of a consistent 
treatment of a Ward identity (vertex and self-energy) without the need of
confinement.

\subsection{The Solution}

In the previous section we have neglected the effects due to the asymmetry 
in $n_{\pm}$ as they appear to be subleading in $1/N$, along with the quark
loops. In a way, we derived a quenched result. However, the occurrence of 
strong infrared sensitive terms of the form $x/m\sim 1/\sqrt{N}m$ in the 
diagrammatic expansion requires that we resum them, prior to taking 
$N\rightarrow\infty$ for finite $m$. We show below, that the infrared
sensitive terms upset naive power counting in the axial singlet
channel. They are just a manifestation of a screening phenomenon in a 
statistical ensemble made of negative and positive topological charges.
Technically the necessity of rederiving $1/N$ counting rules stems from the 
fact that we have introduced here a non-large $N$ ingredient --- the $\chi$ 
field.

The new diagrams that can contribute to our above analysis
are the formerly disconnected quark loops which now interact via 
$\chi$ propagators. 
Looking at a few diagrams we conclude that

\begin{itemize}
\item the $\chi$ lines cannot end on the same quark loop. This gives
 a subleading contribution in $1/N$.
\item when considering the Green's function ( a 1-point correlator ) we
 get no additional graphs.
\item two point correlators of the type
\eq
\cor{\psi^{\dagger}\Gamma\psi \psi^{\dagger} \Gamma \psi}
\eqx
get contributions on the endpoints from graphs proportional to $x$ interacting
through a dressed $\chi$ propagator.
\item the dressed $\chi$ propagator gets contributions from the part of a 
closed quark loop which is proportional to $x^2$.
\end{itemize}

\noindent With the above observations, we now proceed to solve the puzzle.

\subsection{Screening}

Here we isolate the contribution of the closed quark loop $\Gamma$ which is
of order $\chi^2$. For that consider a general Feynman graph made up 
of {\em undressed} quark and
gluon propagators.  A factor proportional to $\chi$ can only appear
from a summation in a quark loop (with the $\gmv$ part of a propagator
contributing). Therefore we must have two distinguished quark loops in
the graph $\Gamma$. Now we may reduce the parts of $\Gamma$ lying `at
the edges' and between the distinguished loops. The former transform
into a trace of the (symmetric) Green's function, while the latter one
becomes the resummed gluon propagator (\ref{e.resum}). We must yet
ensure that each loop picks up a factor of $\chi$.
For this we must take only the $\gmv$ channel of the resummed
propagator. In the end we obtain:
\eq
-(-1)\tr(\GG\gmv\lm_I)\cdot
\f{1}{2N} \left[\f{1}{2-d}\right]_{IJ}
\tr(\GG\gmv\lm_J)
\eqx
The first minus sign comes from the fermion loop, while the second one
from the resummed gluon propagator.
This gives (recall eq. (\ref{e.green}))
\eq
-\chi^2 c_I \f{1}{2N} \left[\f{1}{2-d}\right]_{IJ} c_J
\eqx
with $c_I= i{\rm tr}\, \GG \lm_I$. Resummation gives
\eq
\chi_{top}=\f{1}{\f{1}{\chis}+\f{1}{2} 
c_I \left[\f{1}{2-d}\right]_{IJ} c_J}
\eqx
Using the definitions for $c_I$ and $d_{IJ}$ we obtain
\eqn
\f{1}{\chi_{top}}-\f{1}{\chis}&=&
\sum_{i=1}^{N_f}\f{1}{m_i\sqrt{4+m_i^2}+m_i^2}
\label{MAIN1}
\eqnx
For $N_f=0$ (quenched) the topological susceptibility is $\chi_{top}=\chi_*$,
as it should. For $N_f\neq 0$ (unquenched) the topological susceptibility 
is screened by the quark loops and vanishes in the chiral limit 
\cite{SHIFMAN}. The alternative, non-diagrammatic proof of equality
(\ref{MAIN1}) is presented in Appendix~A.

\subsection{Pseudoscalar susceptibility}

\begin{figure}[htbp]
\setlength{\unitlength}{1mm}
\centerline{
\parbox{30mm}{%
  \begin{fmfgraph*}(30,7.5)
	\fmftop{i,v1,vh1,vh2,v2,o}
	\fmfforce{(0,.5h)}{i}
	\fmfforce{(.125w,h)}{v1}
	\fmfforce{(.25w,.5h)}{vh1}
	\fmfforce{(.75w,.5h)}{vh2}
	\fmfforce{(.875w,h)}{v2}
	\fmfforce{(w,.5h)}{o}
	\fmf{plain,left=1}{i,vh1,i}
	\fmf{plain,left=1}{o,vh2,o}
	\fmf{dashes,label=dressed,l.si=right}{vh1,vh2}
	\fmfv{d.sh=circle,d.f=1,d.si=.07w}{v1}
	\fmfv{d.sh=circle,d.f=1,d.si=.07w}{v2}
	\fmfv{d.sh=polycross 4,d.si=.1w,d.f=0}{i}
	\fmfv{d.sh=polycross 4,d.si=.1w,d.f=0}{o}
  \end{fmfgraph*}
}}
\vspace*{3mm}
\caption{`Trace-trace' contribution ($\Gamma_3$) to the pseudoscalar
susceptibility.}
\label{fig-pss}
\end{figure}
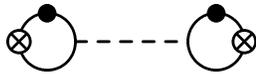

Given the above observations, we conclude that our previous calculation
for $\Gamma_1$ and $\Gamma_2$ is correct. However our neglect of the 
term from the second line in~(\ref{TRTR})
is not correct. Although, this term cannot receive a contribution from
gluon insertions as they are suppressed in $1/N$, they may have a contribution
through a fluctuation in the size of the quark matrices ${\bf Q}$. This is 
essentially an exchange of a $\chi\sim \sqrt{N}$ field as shown in 
Fig.~\ref{fig-pss}.

The quark loop with $\gmv\lm_0$ insertion gives
\eq
\tr\,\GG\gmv\lm_0=2N\sqrt{\f{2}{3}}\left(\sum_{j=0}^{N_f}\f{i}{m_j}x
+O(x^2)\right)
\eqx
so the relevant part is just $\sqrt{2/3}\sum_j\f{i}{m_j}\chi$. 
The full graph gives
\eq
\Gamma_3=-\f{2}{3}\left(\sum_{i=1}^{N_f}\f{1}{m_i}\right)^2
\f{1}{\f{1}{\chis}+ \sum_{i=1}^{N_f}\f{1}{m_i(\sqrt{4+m_i^2}+m_i)}}
\eqx
The susceptibility is then the sum of the two graphs of
Fig.~\ref{fig-ps} and the graph of Fig.~\ref{fig-pss}, that is
\eq \hspace*{-2mm}
\Gamma_1\!\!+\!\Gamma_2\!-\!\Gamma_3=\!\f{2N}{3}\!\!\left[\left(\sum\f{1}{m_i}
\right)^2\!\!\f{1}{\f{1}{\chis}\!+\!\!\sum S_i}-\!4\sum_i S_i\right]
\eqx
where we used the notation
\eq
S_i=\f{1}{m_i(\sqrt{4+m_i^2}+m_i)}
\eqx
For equal quark masses we get 
\eqn
\chi_{ps}&=&-\f{2N}{3}\left[-12S+\f{9/m^2}{\f{1}{\chis}+3S}\right]
\nonumber \\
&=&
-\f{2N}{3}\f{\f{9}{m^2}-\f{12S}{\chis}-4\cdot 9 S^2}{\f{1}{\chis}+3S}
\label{MAIN2}
\eqnx
Since our choice of $\lambda_0=\sqrt{2/3}$ and $\gamma_5$ is $2N\times 2N$
valued with $n_++n_-=2N$ fixed, the prefactor of $2N/3$ is natural in 
(\ref{MAIN2}). The properly normalized susceptibility is 
\be
\tilde{\chi}_{ps} = \frac 3{2N} \chi_{ps} =\frac 3{2N}
(\Gamma_3-\Gamma_1-\Gamma_2)
\ee
which is finite in the large $N$ limit.

In the chiral limit $S\sim 1/2m$ and the leading singularities in $1/m^2$
cancel in the numerator. The pseudoscalar susceptibility 
$\chi_{ps}$ is finite in the chiral limit, thereby
solving the $U(1)$ problem. The {\it a priori} Goldstone pole in 
$\Gamma_1+\Gamma_2$ was removed (screened) by the order $\sqrt{N}$ 
fluctuations in the size of the matrices. Although our result was derived 
with a gaussian weight in $A$, we believe the latter to be a fixed point
as discussed in \cite{BREZINZEE}.

\subsection{Scalar Susceptibility}

Similar arguments for the scalar susceptibility
\eqn
\chi_{s}=\f{1}{V_4}\int d^4x\cor{T^*\psib\psi(x)\,\psib\psi(0)}_c
\label{scalarsus}
\eqnx
translates in ChRMM to
\eqn
\chi_{s}&=&\f{1}{N}\left[-\cor{\tr {\bf Q}^{-1} {\bf Q}^{-1}} \right] 
\nonumber \\
 &+&\f{1}{N}\left[\cor{\tr ({\bf Q}^{-1}
) \tr({\bf Q}^{-1})}_c\right]
\label{TRr}
\eqnx
ignoring (ultraviolet sensitive) contributions from continuum states.
Similar correlators using also results from random matrix theory where 
also discussed in \cite{USLONG}.

Since the scalar susceptibility is proportional to the derivative of the
Green's function with respect to the mass, we can immediately 
write down the result for (\ref{TRr}) in the form
\eqn
\chi_{s}=N_f\left(1-\frac{m}{\sqrt{4+m^2}}\right)
\label{MAIN3}
\eqnx 
This result is only qualitative, since we have ignored the effects of the scale 
anomaly. Also for small values of $m$, (\ref{MAIN3}) does not show terms of the 
form ${\rm ln} m$ expected from the exchange of two Goldstone modes. In ChRMM
these exchanges are $1/N$ suppressed.

\subsection{Numerical Results}

\begin{figure}[htbp]
\centerline{\epsfxsize=8.5cm \epsfbox{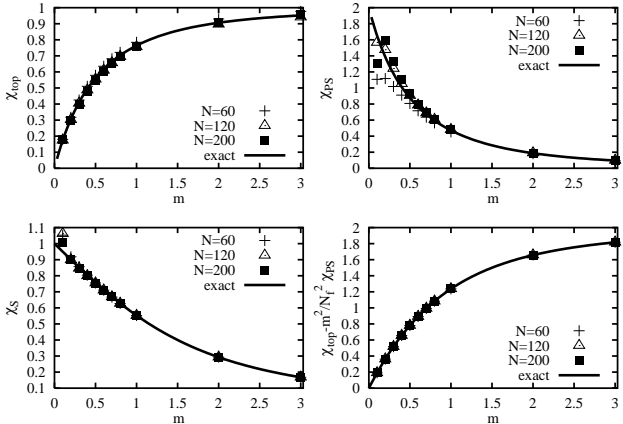}}
\caption{Normalized topological (upper left), pseudoscalar 
(upper right), scalar (lower left) susceptibilities, and
Ward identity (lower right)  for $N_f=1$. The numerical simulations where
carried for fixed $2N=n_+ +n_-=  60, 120, 200$ and 
$<(n_+-n_-)^2>=N\chi_{\star}=N$. The solid line is our analytical result.}
\label{sus1fla}
\end{figure}

\begin{figure}[htbp]
\centerline{\epsfxsize=8.5cm \epsfbox{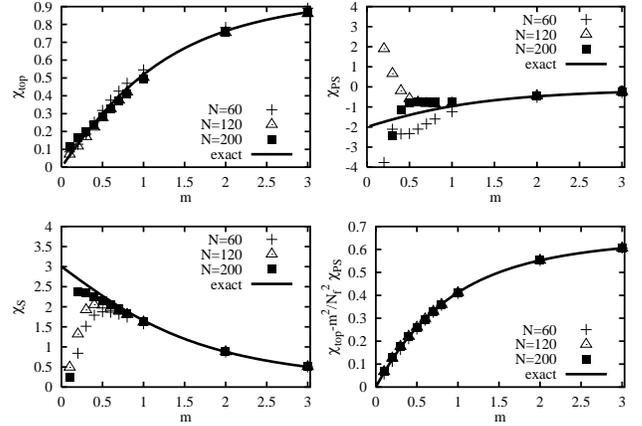}}
\caption{Same as Fig.~\ref{sus1fla} but for $N_f=3$.}
\label{sus3fla}
\end{figure}

To check on the validity of the above arguments, we have carried out direct
numerical calculations using a gaussian distributed ensemble of rectangular
matrices. Figures~\ref{sus1fla} and ~\ref{sus3fla} show the results 
of the numerical simulations for the topological, pseudoscalar and scalar 
susceptibilities, with one and three flavors respectively. The solid
lines follow from (\ref{MAIN1}), (\ref{MAIN2}), and (\ref{MAIN3}) respectively.
Our $1/N$ analysis is confirmed, except for very small values of $m$,
where finite $1/N$ size effects are noted.
Throughout $\chi_*$ was 
set to 1.

We have checked that our numerical results are in agreement with the Ward
identity (\ref{6}) when translated to ChRMM. Specifically
\eqn
\chi_{top} +\frac{m^2}{N_f^2}\tilde{\chi}_{ps}=-\frac{2mi}{N^2_f}
{\rm tr} {\cal G}
\label{wardrmm}
\eqnx
The proof of the above identity is included in Appendix~B. 
Figures~\ref{sus1fla} and ~\ref{sus3fla} (lower right) show the 
Ward identity (\ref{wardrmm}). The numerical points represent the l.h.s.
of the Ward identity, the solid line is calculated from the analytical
prediction for the r.h.s. using (\ref{SPEC1}) with $x=0$.

\section{Conclusions}

We have shown that an ensemble of asymmetric but standard ChRMM with 
an assumed gaussian distribution for the asymmetry, yields a screened
topological susceptibility and solves the $U(1)$ problem. We believe
this result to be new in the context of ChRMM. Using diagrammatic 
techniques in the context of the $1/N$ approximation, we have shown
the fermion determinant contributes importantly in the axial-singlet
channel through infrared sensitive terms, but otherwise is quenched
in most other channels in the thermodynamical limit. We also believe
that this result is new in the context of ChRMM. Some of these
results are generic and illustrative of the mechanisms at work in 
more realistic models of the QCD vacuum, such as the 
instanton liquid model. It would be interesting to see how the present
arguments extend to nonstandard ChRMM \cite{USNJL}, and how they are 
modified at high temperature and finite $N$ in comparison to current
lattice simulations \cite{DETAR}. Also, our analysis allows for an
assessment of the $\theta$ vacua in ChRMM, as we will discuss next.

\end{fmffile}

\vglue 0.6cm
{\bf \noindent  Acknowledgments \hfil}
\vglue 0.4cm
This work was supported in part  by the US DOE grant DE-FG-88ER40388, by
the NSF grant NSF-PHY-94-21309,
by the Polish Government Project (KBN) grants 2P03B19609, 2P03B08308 and by 
the Hungarian Research Foundation OTKA. One of us (R.A.J)
thanks the Nuclear Theory Group in Stony Brook and GSI for their
hospitality.


\vglue 0.6cm
{\bf \noindent Appendices \hfil}
\vglue 0.4cm

\noindent
{\it A. Screened Topological Susceptibility}\\

In this Appendix, we will outline an alternative derivation in favor of
the screening of the topological charge in the present model.
The connected `vacuum' diagrams (prior to averaging over
$\chi$) are  given by $\log Z(\chi)$. For a fixed asymmetry 
$x=\chi/2N=(n_+-n_-)/2N$ we have 
\eqn
&&\partial_m\log Z(\chi)=-i\la\psi^{\dagger}\psi\ra=
-2N i\, {\rm tr}\,{\cal G} \nonumber \\ &&=
N N_f\left(-m+\sqrt{4+m^2+\f{4x^2}{m^2}}\right)
\eqnx
The contribution of order $x^2$ follows by expanding the right hand side
\eq
(\partial_m\log Z(\chi))_2 =N N_f \f{2x^2}{m^2\sqrt{4+m^2}}
\eqx
and integrate it with respect to $m$ to obtain
\eq
(\log Z(\chi))_2 = N N_f \left(-\f{\sqrt{m^2+4}}{2m} +const\right)x^2 
\eqx
The constant is set by the requirement that for $m\rightarrow \infty$
we should obtain zero. Hence
\eqn
(\log Z(\chi))_2 &=& N N_f \left(-\f{\sqrt{m^2+4}}{2m} +\f{1}{2}\right)
\cdot x^2=\\ 
&=& -2N N_f \f{1}{m(\sqrt{4+m^2}+m)} \cdot x^2\\
&=& -\f{1}{2N} N_f \f{1}{m(\sqrt{4+m^2}+m)} \cdot \chi^2
\label{APPEN1}
\eqnx
The full partition function is is given by
\eq
Z=\int d\chi e^{-\f{\chi^2}{2\chi_* N}} Z(\chi).
\label{APPEN2}
\eqx
Inserting (\ref{APPEN1}) into (\ref{APPEN2}) yields
\eq
Z=\int d\chi e^{-\f{1}{2N}\left(\f{1}{\chi_*}+N_f
\f{1}{m(\sqrt{4+m^2}+m)} \right) \chi^2+\ldots}
\eqx
in agreement with (\ref{MAIN1}).

\newpage 
\noindent
{\it B. Ward identity in ChRMM}\\

The Ward identity (\ref{wardrmm}) can be derived from the partition function 
for finite $\theta$ angle and massive quarks. Specifically
\eq
Z[\th]=\la\mbox{det}^{N_f}
\left( \begin{array}{cc}
  ime^{i\f{\th}{N_f}} \!\!\!\! & A^{\dagger} \cr
  A & \!\!\!\! ime^{-i\f{\th}{N_f}} \end{array}\right) \ra=
\la\det{}^{N_f} {\bf Q}_\th\ra
\eqx
Using the derivatives at $\th=0$ 
\eq
\partial_\th {\bf Q}_\th=-\f{m}{N_f}\gmv
\eqx
and
\eq
\partial^2_\th {\bf Q}_\th=-\f{im}{N_f^2}\One
\eqx
we get
\eq
N\chi_{top}=\f{-im}{N_f^2}\la\psi^{\dagger}\psi\ra
-\f{m^2}{N_f^2}\la\psi^{\dagger}
\gmv\psi
\psi^{\dagger}\gmv\psi\ra
\eqx
Since $\la\psi^{\dagger}\psi\ra=2N {\rm tr}\, {\cal G}$, then
\eq
\chi_{top}=-\f{2m}{N_f^2} i\, {\rm tr }\, \GG-\f{m^2}{N_f^2}\tilde\chi_{ps}
\eqx
in agreement with (\ref{wardrmm}).


\end{document}